\documentclass[preprint,pra,aps,showpacs,preprintnumbers,amsmath,amssymb,eqsecnum]{revtex4}
\usepackage{graphicx}% Include figure files
\usepackage{dcolumn}% Align table columns on decimal point
\usepackage{bm}% bold math
\usepackage{verbatim}
\usepackage{epsfig}
\usepackage{epstopdf}

\def\eps{{\epsilon}}
\def\z{{\mathbf{z}}}
\def\hP{{ \hat P}}
\def\hp{{ \hat p}}
\begin{document}

%\preprint{Imperial/TP/***}

\title{Pitfalls of Path Integrals: Amplitudes for Spacetime Regions and the Quantum
Zeno Effect}

\author{J.J.Halliwell}%
\email{j.halliwell@imperial.ac.uk}

\affiliation{Blackett Laboratory \\ Imperial College \\ London SW7
2BZ \\ UK }

\author{J.M.Yearsley}
\email{jmy27@cam.ac.uk}
\affiliation{Centre for Quantum Information and Foundations, DAMTP, Centre for Mathematical Sciences, University of Cambridge, Wilberforce Road, Cambridge CB3 0WA, UK}

%\affiliation{


\begin{abstract}
Path integrals appear to offer natural and intuitively appealing methods for defining quantum-mechanical amplitudes for questions involving spacetime regions. For example, the amplitude for entering a spatial region during a given time interval is typically defined by summing over all paths between given initial and final points but restricting them to pass through the region at any time. We argue that there is, however, under very general conditions, a significant complication in such constructions. This is the fact that the concrete implementation of the restrictions on paths over an interval of time corresponds, in an operator language, to sharp monitoring at every moment of time in the given time interval. Such processes suffer from the quantum Zeno effect -- the continual monitoring of a quantum system in a Hilbert subspace prevents its state from leaving that subspace.
As a consequence, path integral amplitudes defined in this seemingly obvious way have physically and intuitively unreasonable properties and in particular, no sensible classical limit. In this paper we describe this frequently-occurring but little-appreciated phenomenon in some detail, showing clearly the connection with the quantum Zeno effect. We then show that it may be avoided by implementing the restriction on paths in the path integral in a ``softer'' way. The resulting amplitudes then involve a new coarse graining parameter, which may be taken to be a timescale $\eps$, describing the softening of the restrictions on the paths. We argue that the complications arising from the Zeno effect are then negligible as long as $\eps \gg 1/ E$, where $E$ is the energy scale of the incoming state.

\end{abstract}


%\pacs{03.65.Xp, 03.65.Yz., 03.65.Ta}

%\centerline{Imperial/TP/03-4/7}

\maketitle

\newcommand\beq{\begin{equation}}
\newcommand\eeq{\end{equation}}
\newcommand\bea{\begin{eqnarray}}
\newcommand\eea{\end{eqnarray}}

\def\A{{\cal A}}
\def\D{\Delta}
\def\H{{\cal H}}
\def\E{{\cal E}}
\def\p{\partial}
\def\la{\langle}
\def\ra{\rangle}
\def\ria{\rightarrow}
\def\x{{\bf x}}
\def\y{{\bf y}}
\def\k{{\bf k}}
\def\q{{\bf q}}
\def\p{{\bf p}}
\def\P{{\bf P}}
\def\r{{\bf r}}
\def\s{{\sigma}}
\def\a{\alpha}
\def\b{\beta}
\def\e{\epsilon}
\def\U{\Upsilon}
\def\G{\Gamma}
\def\om{{\omega}}
\def\Tr{{\rm Tr}}
\def\ih{{ \frac {i} { \hbar} }}
\def\trho{{\rho}}

\def\au{{\underline \alpha}}
\def\bu{{\underline \beta}}
\def\pp{{\prime\prime}}
\def\id{{1 \!\! 1 }}
\def\half{\frac {1} {2}}

\def\eps{{\epsilon}}
\def\z{{\mathbf{z}}}
\def\hP{{ \hat P}}
\def\hp{{ \hat p}}
\def\x0{{{\bf x}_0}}
\def\xf{{{\bf x}_f}}

\def\jjh{j.halliwell@ic.ac.uk}

\section{Introduction}
Consider the following question in non-relativistic quantum mechanics for a point particle in $d$ dimensions:
What is the amplitude $g_{\Delta} (\xf, t_f | \x0, t_0 )$ for the particle to start at a spacetime
point $(\x0, t_0)$, pass through a spatial region $\Delta$ and end at a spacetime point $ (\xf, t_f) $? The seemingly-obvious answer to this question is surely the path integral expression,
\beq
g_{\Delta} (\xf, t_f | \x0, t_0 ) = \int_{\Delta}
{ \mathcal D} {\bf x} (t) \ \exp \left( i \int_{t_0}^{t_f} dt \left[ \half m \dot {\bf x}^2 - U( {\bf x} ) \right] \right)
\label{1.1}
\eeq
(We choose units in which $\hbar=1$).
In this expression, the paths ${\bf x} (t) $ summed over satisfy the initial condition ${\bf x} (t_0) = \x0$, the final condition ${\bf x} (t_f) =\xf $ and pass, at any
intermediate time, through the region $\Delta$ \cite{Fey,FeHi,Sch}, as depicted in Fig.1.
\begin{figure}[htbp]
   \centering
   \includegraphics[width=4in]{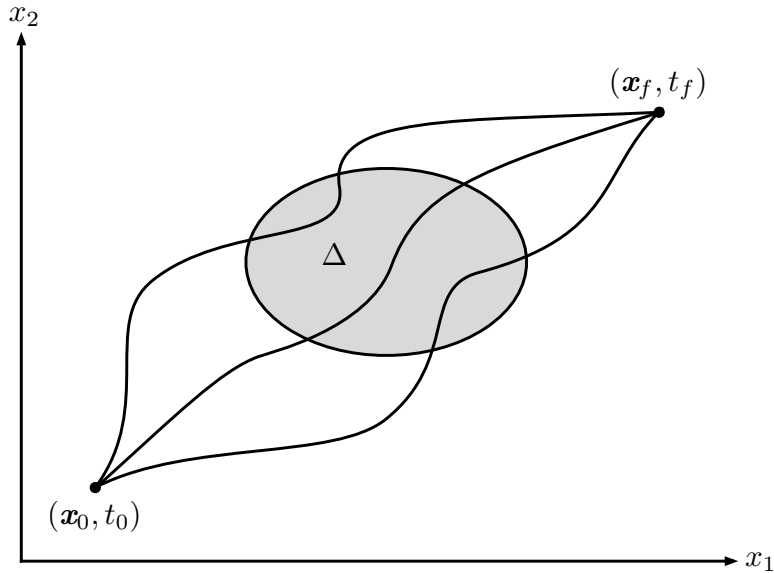}
   \caption{The amplitude Eq.(\ref{1.1}) is obtained by summing over paths which enter the spatial region $\Delta$ at any time between $t_0$ and $t_f$.}
   \label{Fig1}
\end{figure}
The point of this paper is to argue that there are potential problems with this seemingly-obvious answer.

We first develop these ideas further.
We might similarly assert that the amplitude $ g_{r} (\xf, t_f | \x0, t_0 ) $ for the particle never entering the region $\Delta $ is a given by a path integral expression of the form
\beq
g_{r} (\xf, t_f | \x0, t_0 ) = \int_{r}
{ \mathcal D} {\bf x} (t) \ \exp \left( i \int_{t_0}^{t_f} dt \left[ \half m \dot {\bf x}^2 - U( {\bf x} ) \right] \right)
\label{1.2}
\eeq
This object, the restricted propagator, is given by
a sum over paths restricted to lie always outside $\Delta$. We then clearly have
\beq
g (\xf, t_f | \x0, t_0 ) =
g_{\Delta} (\xf, t_f | \x0, t_0 )
+ g_{r} (\xf, t_f | \x0, t_0 )
\label{1.3}
\eeq
where $g$ is the usual propagator obtained by summing over all paths from initial to final point.

Objects such as Eq.(\ref{1.2}) have to be given proper mathematical
definition. For the moment, we have in mind a definition involving the usual time-slicing
procedure, in which the time interval is divided up into $n$ equal intervals of size $\epsilon$ so that $ t_f - t_0 = n \epsilon $ and we consider propagation between
slices labelled by times $t_k = t_0 + k \eps $, where $k=0,1 \cdots n$. Eq.(\ref{1.2})
is then defined by a limit of the form
\beq
g_{r} (\xf, t_f | \x0, t_0 ) = \lim_{\epsilon \rightarrow 0, n \rightarrow \infty}
\int_r d^d x_1 \cdots \int_r d^d x_{n-1}
\prod_{k=1}^{n}  g ( {\bf x}_{k}, t_{k} | {\bf x}_{k-1}, t_{k-1} )
\label{1.4a}
\eeq
where ${\bf x}_n = \xf$ and the integrals are over the region outside $\Delta$.
The propagator is approximated by
\beq
g ( {\bf x}_{k}, t_{k} | {\bf x}_{k-1}, t_{k-1} ) = \left( \frac {m}{2 \pi i \eps}
\right)^{d/2} \exp \left( i S ( {\bf x}_{k}, t_{k} | {\bf x}_{k-1}, t_{k-1} ) \right)
\eeq
for small times, where the exponent is the action between the indicated initial and final
points and the limit is taken in such a way that $n \eps$ is fixed.

Writing the above amplitudes in operator form, $\hat g_{\Delta}$, $\hat g_r $,
where, for example,
\beq
g_{r} (\xf, t_f | \x0, t_0 ) = \langle \xf | \hat g_{r} (t_f,t_0) |
\x0 \rangle
\label{1.4}
\eeq
%and we use $\hat g(t_f,t_0) $ to denote $ \exp (- i H (t_f -t_0) ) $,
it seems reasonable to suppose that the probabilities for a particle in initial state $|\psi \rangle $ entering or not entering the region $\Delta$ during the time interval
$[t_0,t_f]$ are given by
\bea
p_{\Delta} &=& \langle \psi | \hat g_{\Delta} (t_f,t_0)^\dag  \hat g_{\Delta} (t_f,t_0) | \psi \rangle
\label{1.5}
\\
p_{r} &=& \langle \psi | \hat g_{r} (t_f,t_0)^\dag  \hat g_{r} (t_f,t_0) | \psi \rangle
\label{1.6}
\eea
To be sensible probabilities these expressions should obey the simple sum rule
\beq
p_\Delta + p_r = 1
\label{1.7}
\eeq
but using $\hat g = \hat g_\Delta + \hat g_r $, it is easy to see that this is generally not the case
unless the interference between the two types of paths vanishes, which means that
\beq
{\rm Re} \ \langle \psi | \hat g_{r} (t_f,t_0)^\dag \hat g_\Delta (t_f,t_0) | \psi \rangle = 0
\label{1.8}
\eeq
This condition can hold, perhaps approximately, for certain types of initial states, although this may be non-trivial to prove and there is no guarantee that those states are physically interesting ones.
%Using this procedure one might attempt to compute the probability that an incoming %wave packet enters the region $\Delta$ at any time during the time interval $[t_0,t_f]$.

Path integral expressions of the above general type have been postulated and used extensively in a wide variety of circumstances in which time is involved in a non-trivial way.
Indeed, Feynman's original paper on path integrals was entitled,
``Space-time approach to non-relativistic quantum mechanics'', clearly suggesting that the spacetime features of the path integral construction should be made use of \cite{Fey}.
Path integral constructions have for many years played an important role in
quantum cosmology and quantum gravity \cite{Har0,HaHa1,FRW,QC1,QC,QC2,Har3,HaMa,HaTh1,HaTh2,HaWa,Hal,CrHa,CrSi,Whe,
ChWa,AnSa1,AnSa2,Schr}.  They have also been particularly useful
in addressing issues concerning time in non-relativistic quantum mechanics, for example
in studies of the arrival time \cite{YaT,Yam2,Yam3,Har4,MiHa,HaZa,HaYe1,HaYe2,Ye0,AnSa3,AnSa4,AnSa5}, dwell time  and tunneling time \cite{Fer,Yam0,Yam1,GVY,Sok}. Path integrals are also useful ways of formulating
continuous quantum measurement theory \cite{Caves,Mensky}

Many of these applications of path integrals are in the specific framework of the decoherent histories approach to quantum theory \cite{GeH1,GeH2,Gri,Omn,Hal2,DoH,Hal5,Ish,Ish2,IsLi,ILSS}
(in which the relations Eqs.(\ref{1.7}), (\ref{1.8}) arise) and quantum measure theory \cite{Sor,Sal}. Also related are the various attempts to derive the Hilbert space formulation of quantum theory from path integral
constructions \cite{Dow}.
Here we wish to study path integral expressions as entities in their own right without being tied to a specific approach to quantum theory.

However, as indicated, there is a problem with the innocent-looking and informally appealing path integral expressions Eqs.(\ref{1.1}), (\ref{1.2}), with the consequence that their properties can be very different to intuitive expectations. This difficulty goes beyond the problem of satisfying the no-interference condition, Eq.(\ref{1.8}), although is related to it. To see it, let us focus on
the amplitude for not entering the region $\Delta$, Eq.(\ref{1.2}).
The key issue is that the innocent-looking restriction on paths effectively means that an initial state propagated by Eq.(\ref{1.2}) is required to have zero support in the region $\Delta$ at {\it every moment of time} between $t_1$ and $t_2$. Evolution with this propagator therefore suffers from
the quantum Zeno effect -- the fact that continual monitoring of a quantum system in a Hilbert subspace prevents it from leaving that subspace \cite{Kha,Zeno,CSM,Peres,Kraus,Sud,Sch2,Wall,Wall2}.
The consequence is that the amplitude Eq.(\ref{1.2}), or equivalently $\hat g_r (t_2,t_1)$, actually describes {\it unitary} propagation on the Hilbert space of states with support only in the region outside $\Delta$ and therefore gives probability $ p_r= 1$ for any incoming state. This then means that either the sum rule Eq.(\ref{1.7}) is not satisfied in which case the probabilities are not meaningful, or that it is satisfied but $ p_{\Delta} = 0$, which means that any incoming state aimed at $\Delta$ has probability zero for entering that region, a physically nonsensical result.

Differently put, the innocent-looking restriction on paths in the restricted propagator Eq.(\ref{1.2}) with the usual implementation Eq.(\ref{1.4a})
effectively sets {\it reflecting} boundary conditions on the boundary of $\Delta$, which means that any incoming state is totally reflected under propagation  by $ \hat g_r $. To obtain the intuitively sensible result, we would need a propagator analogous to Eq.(\ref{1.2}) in which the incoming state is {\it absorbed}.

One could perhaps look for another way out, which is to suppose that
$\hat g_\Delta$ is related to some sort of measurement scheme, in which case there
is no obligation to satisfy the probability sum rule and that Eq.(\ref{1.5}) may
then give a reasonable formula for the probability of entering. To this end it is
useful to give a more detailed formula for $g_{\Delta}$ using the path decomposition
expansion (PDX) \cite{PDX,HaOr,Hal3}. The paths summed over in Eq.(\ref{1.1}) may be partitioned according
to the time $t$ and location ${\bf y}$ at which they cross the boundary $\Sigma$
of $\Delta$ for the first time. (See Fig.2)
\begin{figure}[htbp]
   \centering
   \includegraphics[width=4in]{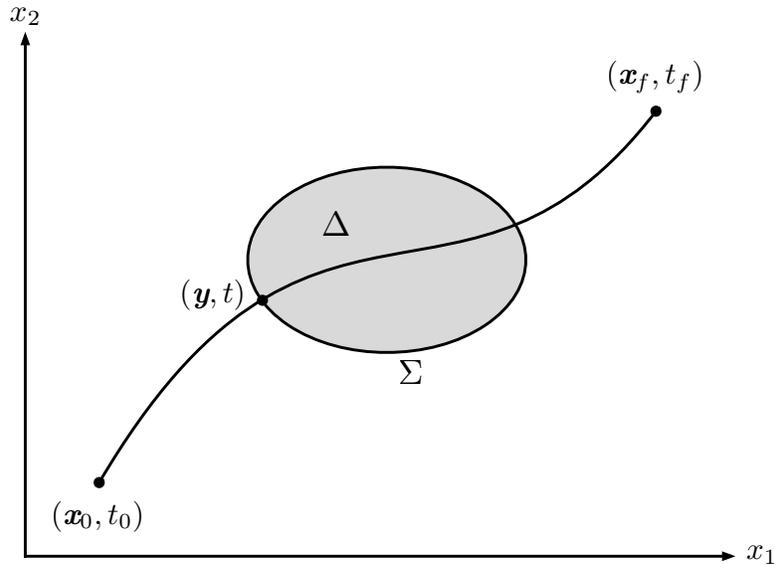}
   \caption{The path decomposition expansion Eq.(\ref{PDX1}). Each path from $({\bf
   x_0,t_0)}$ to $({\bf x}_f,t_f)$ passing through $\Delta$ may be labeled by its time $t$ and location ${\bf y}$ of first crossing of the boundary $\Sigma$.}
   \label{Fig2}
\end{figure}
As a consequence, it is possible to derive the formula
\beq
g_{\Delta} (\xf, t_f | \x0, t_0 ) = \frac{i} {2m} \int_{t_0}^{t_f} dt
\int_{\Sigma} d^{d-1} y
\ g ( \xf, t_f | {\bf y}, t ) \ {\bf n} \cdot \nabla_{\bf x}
g_r ({\bf x}, t | \x0, t_0 ) \big|_{{\bf x} = {\bf y}}
\label{PDX1}
\eeq
where ${\bf n}$ is the outward pointing normal to $\Sigma$. The restricted
propagator $g_r({\bf x}, t | {\bf x_1}, t_0 ) $ vanishes when either end is on $\Sigma$
but its normal derivative does not. In fact, the normal derivative of the restricted
propagator in Eq.(\ref{PDX1}) represents the sum over paths which are restricted to remain outside $\Delta$ but end on its boundary \cite{PDX,HaOr,Hal3}.

Although on the face of it, Eq.(\ref{PDX1}) is a plausible formula for the crossing amplitude, the problem
with this expression, which is fully equivalent to Eq.(\ref{1.3}), concerns the treatment
of paths that are outside $\Delta$ before their first crossing at $t$. The fact that
these paths are represented using the restricted propagator means that any such paths
arriving at $\Sigma$ before $t$ are reflected, rather than simply dropped from the
sum. This means an incoming wave packet propagated using Eq.(\ref{PDX1}) will be partly reflected
and we anticipate that this formula could fail to give intuitively
sensible results.
%[NOTE: Not sure about this bit but I think it is key].

One way or another, we find that the simple and seemingly obvious notion of ``restricting paths" leads to the quantum Zeno effect and hence to unphysical results. This is the sense in which we would say that path integral constructions may suffer from pitfalls.

The purpose of this paper is to give a concise statement of the general problem outlined above with naive path integral expressions of the form Eqs.(\ref{1.1}), (\ref{1.2}), and to point out how to construct practically useful modified path integrals which give a meaningful answer to the question of assigning amplitudes to spacetime regions with a sensible classical limit.

Many papers using path integral constructions of the above type appear to be oblivious to this effect and its consequences. Although the results of such papers are not necessarily wrong, they often do not have sensible classical limits and in measurement-based models, they may, unknowingly, only be valid in the regime of strong measurement.
These problems with the Zeno effect in path integral constructions can be seen in
a number of early attempts to define probabilities for spacetime regions in the context
of the decoherent histories approach \cite{YaT,Yam2,Yam3,Har4,MiHa,HaZa,AnSa3} and in some
papers on the decoherent histories approach to quantum cosmology \cite{Har3,HaMa,HaTh1,HaTh2,HaWa}.
A possible resolution was given in Refs.\cite{HaYe1,Hal}. There are undoubtedly many
other places in which this difficulty has been encountered.

Here, our aim is to give a general account of the problem and solution, independent of specific approaches to quantum theory and of specific applications. We give a detailed formulation of
the problem in Section 2, the proposed solution in Section 3 and an example involving
the arrival time in Section 4. In Section 5, we give a detailed discussion of the
connections with other work. We summarize and conclude in Section 6.

\section{Detailed Formulation of the Problem}

We first discuss the generality of the problem with path integrals outlined above.
The example in Eqs.(\ref{1.1}), (\ref{1.2}) is in non-relativistic quantum mechanics in
any number of dimensions. The region $\Delta$ can consist of a number of disconnected pieces and then there
could be many different ways of partitioning the paths according to how many different regions they enter or not. We will also assume that $\Delta$ is reasonably large and that its boundary is reasonably smooth, in comparison to any quantum-mechanical lengthscales set by the incoming state.
Furthermore, in the situation depicted in Fig.1, the spatial region is constant in time, but there is no obstruction to allowing it to vary with time. For example, in relativistic quantum mechanics, it may be natural to look at
regions whose boundaries are null surfaces \cite{Hal}.
One could also contemplate path integrals in curved spacetimes which may have unusual properties, such
as closed timelike curves \cite{Har3}, but the basic ideas of path integration are still applicable.
An important area of application of these ideas is to quantum cosmology, where there is no explicit physical time coordinate, but the basic object Eq.(\ref{1.1}) is still the appropriate starting point for the construction of class operators in the decoherent histories analysis of quantum cosmology \cite{Har3,HaWa,Hal}.
This is clearly not an exhaustive list of possibilities, but the arguments presented in what follows will apply to all of these
cases, even though they are presented in the context of the example Eqs.(\ref{1.1}), (\ref{1.2}).

We now set out in more mathematical detail the argument outlined in the Introduction.
We focus on the path integral representation of the restricted propagator Eq.(\ref{1.2}) and its time-slicing representation
Eq.(\ref{1.4a}). We will give an equivalent operator form for Eq.(\ref{1.4a}).
We introduce the projector onto $\Delta$,
\beq
P = \int_{\Delta} d^d x \ | {\bf x} \rangle \langle {\bf x} |
\label{2.2}
\eeq
and its negation, $ \bar P = 1 - P $, the projector onto the region $\bar \Delta$,
the region of $ {\mathbb R}^d$ outside of $\Delta$.
Again we divide the time interval up into $n$ equal intervals of size $\epsilon$ so that
$ t_f - t_0 = n \epsilon $. It is then easy to see that, in terms of the operator $ \hat g_{r}$ defined in Eq.(\ref{1.4}), the time-slicing expression Eq.(\ref{1.4a}) is equivalent to
\beq
\hat g_{r} (t_f,t_0)= \lim_{\epsilon \rightarrow 0, n \rightarrow \infty} \ \bar P e^{ - i H \eps } \bar P \cdots e^{- i H \eps} \bar P
\label{2.3}
\eeq
where there are $n+1$ projectors and $n$ unitary evolution factors $e^{ - i H \eps}$ and
the limit
%$\eps \rightarrow 0 $, $ n \rightarrow \infty$
is taken in such a way that $ n \eps = t_f - t_0 $ is fixed.
That is, inserted in Eq.(\ref{1.4}),
Eq.(\ref{2.3}) gives the time-slicing definition of the path integral expression,
Eq.(\ref{1.4a}).
(Note that there are $n+1$ projectors $\bar P$ in Eq.(\ref{2.3}) but
only $n-1$ corresponding integrals in Eq.(\ref{1.4a}). The two extra projectors
are redundant, and hence the two expressions completely equivalent, if we take
$\x0$ and $\xf$ to be outside $\Delta$).

One can clearly see from the operator form Eq.(\ref{2.3}) that it involves ``monitoring'' of the particle to check if it is in $\bar \Delta$ at each instant of time.
The limit in Eq.(\ref{2.3}) may be computed explicitly \cite{Sch2} and leads to the explicit form
\beq
\hat g_{r} (t_f,t_0) = \bar P \exp \left( - i \bar P H \bar P (t_f - t_0) \right)
\label{2.4}
\eeq
This propagator is, as claimed, unitary on the Hilbert space of states with support in $\bar \Delta$ \cite{Sch2}. It therefore describes the situation in which an incoming state never actually leaves $\bar \Delta$ due to monitoring becoming infinitely frequent, which is clearly the quantum Zeno effect. In simple examples of the restricted propagator, one can easily see that an incoming state is totally reflected off the boundary of the region $\Delta$. Eq.(\ref{2.4}) and its properties explain why the naive path integral expressions Eqs.(\ref{1.1}), ({1.2}) have counter-intuitive properties which lead to unphysical results if not used sufficiently carefully.

It is also reasonable to consider other possible methods for defining the
path integral Eq.(\ref{1.2}). Another natural method is to consider the imaginary
time version of Eq.(\ref{1.2}) and then define the path integral in terms of the
limit of a stochastic process involving random walks on a spacetime lattice (in the
case $U=0$, for simplicity) \cite{Har4,JaGl}. Restricting the paths to lie outside $\Delta$ means finding a suitable boundary condition on the random walks at the boundary of $\Delta$.
Most studies of this problem have imposed reflecting boundary conditions, which appear
to be the easiest ones to impose in practice. As a consequence, this method of defining
the path integral once again leads to the restricted propagator of the form
Eq.(\ref{2.4}). However, it is not clear that one is compelled to use reflecting
boundary conditions in such implementations of the path integral. We will discuss this further below.

\section{Proposed Solution}

Since the Zeno effect is the root of the problem, the solution is clearly to limit
or soften the monitoring of the system in some way in Eq.(\ref{2.3}) so that reflection is minimized. The first obvious way to do this is to decline to take the limit in Eq.(\ref{2.3}) and keep the time-spacing $\eps$ finite.  The second way is to replace
the exact projectors by POVMs. These solutions have been explored in the specific
context of the arrival time problem in one dimension \cite{HaYe1,HaYe2}, but the purpose here is to
present these solutions in a more generally applicable way.

In the first approach,
we therefore define a modified propagator for not entering $\Delta$ by
\beq
\hat g^{\eps}_{r} (t_f,t_0)= \bar P e^{ - i H \eps } \bar P \cdots e^{- i H \eps} \bar P
\label{3.1}
\eeq
where as before there are $n+1$ projectors and $n$ unitary time operators and $ n\eps = t_f - t_0$.
This object can also be represented by a path integral expression of the form Eq.(\ref{1.2}), except that the paths are required to be outside $\Delta$ only at the $n+1$ times $t_0 + k \eps $, where
$ k = 0 \cdots n $, but between these times the paths are unrestricted. This situation is depicted in Fig.3 for a one-dimensional example in which the region $\Delta$ is the interval $[a,b]$.
\begin{figure}[htbp]
   \centering
   \includegraphics[width=4in]{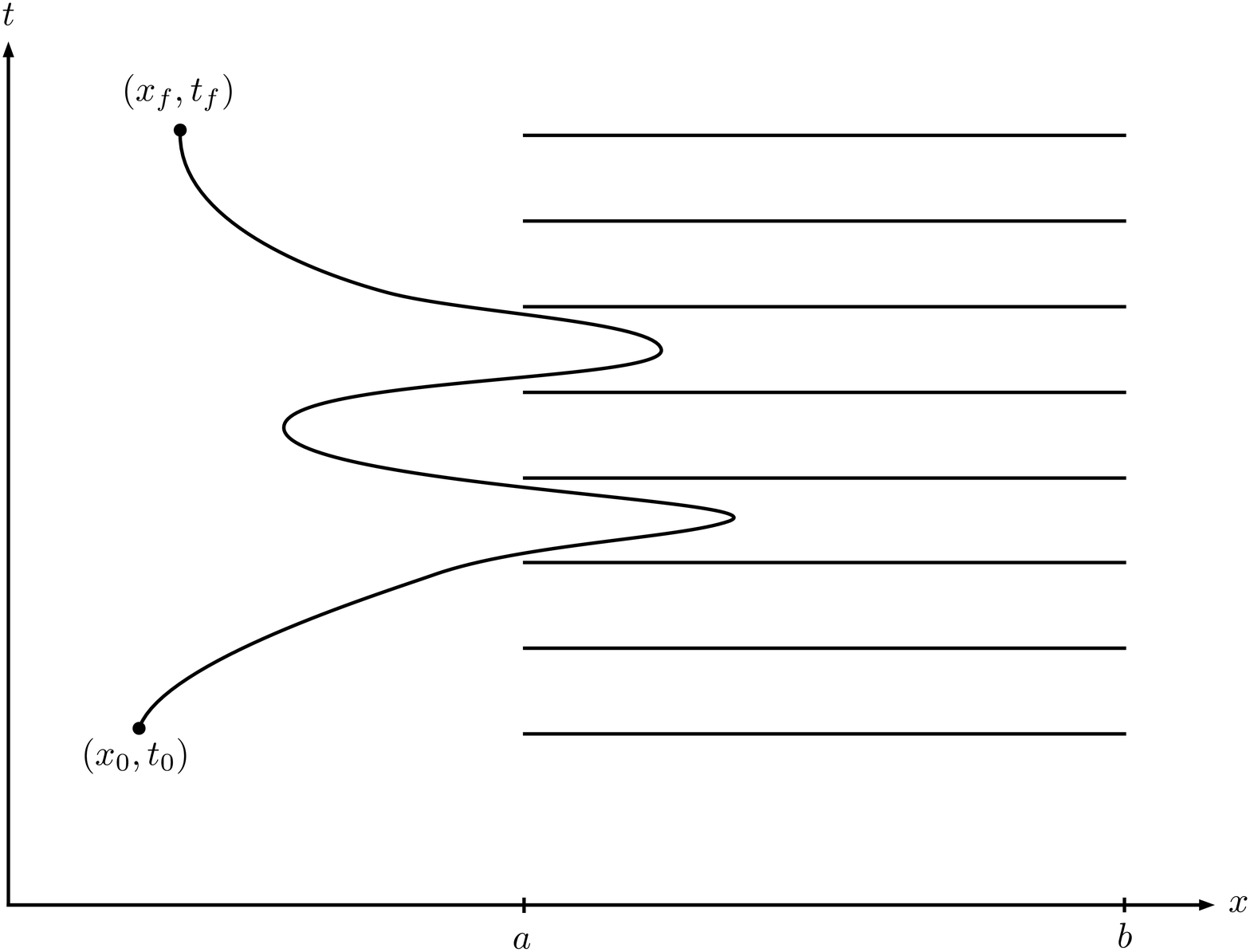}
   \caption{The paths summed over in the path integral representation of the modified propagator Eq.(\ref{3.1}). The paths may enter the region $[a,b]$ but not at the times at which the projectors act. Very wiggly paths, which enter the region frequently, will generally have small contribution to the path integral in a semiclassical approximation, which means that there is an effective suppression of  paths entering the region.}
   \label{Fig3}
\end{figure}

This modified propagator involves a {\it new coarse graining parameter} $ \eps$, describing the precision to within which the paths are monitored. The original propagator Eq.(\ref{1.2}) is obtained in the limit of infinite precision $\eps \rightarrow 0 $. The physically interesting case, however, is that in which $\eps$ is small enough to monitor the paths well, but sufficiently large that an incoming state is not significantly affected by reflection.

There is a very useful approximate alternative to Eq.(\ref{3.1}), which is also helpful in terms of calculating the timescale required to define what ``small'' and ``large'' mean for $\eps$. For the special case of projectors onto the positive $x$-axis in one dimension, $P = \theta (\hat x)$,  it has been shown \cite{Ech,HaYe3} that the string of projectors Eq.(\ref{3.1}) is to a good approximation equivalent to evolution in the presence of a complex potential consisting of a window function on the region
$\Delta$, that is,
\beq
\bar P e^{ - i H \eps } \bar P \cdots e^{- i H \eps} \bar P
\approx \exp \left( - (i H + V_0 P ) (t_f - t_0 ) \right)
\label{3.2}
\eeq
Here, the real parameter $V_0$ depends on $\eps$ and it was shown numerically \cite{HaYe3}
that the best match is obtained with the choice
\beq
\eps V_0 \approx 4/3
\label{3.2a}
\eeq
The approximate equivalence Eq.(\ref{3.2}) is expected to hold when acting on states with energy width $\Delta H$ for which $ \eps \ll 1/ \Delta H $ (a timescale often called the Zeno time). Moreover, the general arguments given in Refs.\cite{Ech,HaYe3}
for the equivalence Eq.(\ref{3.2}) are not obviously tied to the one-dimensional case
with $P = \theta (\hat x)$, so we expect it to hold very generally.

This approximate equivalence means that a second natural candidate for the modified propagator is
\beq
\hat g_r^V (t_f, t_0) =  \exp \left( - (i H + V_0 P ) (t_f - t_0 ) \right)
\label{3.3}
\eeq
As stated, this is approximately equal to Eq.(\ref{3.1}), at least in simple one-dimensional models.
However, it may be taken as an independently-postulated alternative propagator which is essentially equivalent to using the form
Eq.(\ref{2.3}) but replacing the exact projector $\bar P$ with POVMs of the form $\exp( - \eps V_0 P )$, which is our second obvious way of softening the monitoring so as to avoid the Zeno effect.
Moreover, this expression too has a path integral form,
\beq
g^V_{r} (\xf, t_f | \x0, t_0 ) = \int
{ \mathcal D} {\bf x} (t) \ \exp \left( i \int_{t_0}^{t_f} dt \left[ \half m \dot {\bf x}^2 - U( {\bf x})  + i V_0 f_{\Delta} ({\bf x})  \right] \right)
\label{3.4}
\eeq
where $f_{\Delta} ({\bf x})$ is a window function on $\Delta$.
Here, the paths are unrestricted, except at their end points, but paths entering $\Delta$ are suppressed by a complex potential.
This situation is depicted in Fig.4 for the above one-dimensional example, in which the region $\Delta$ is the interval $[a,b]$.
\begin{figure}[htbp]
   \centering
   \includegraphics[width=4in]{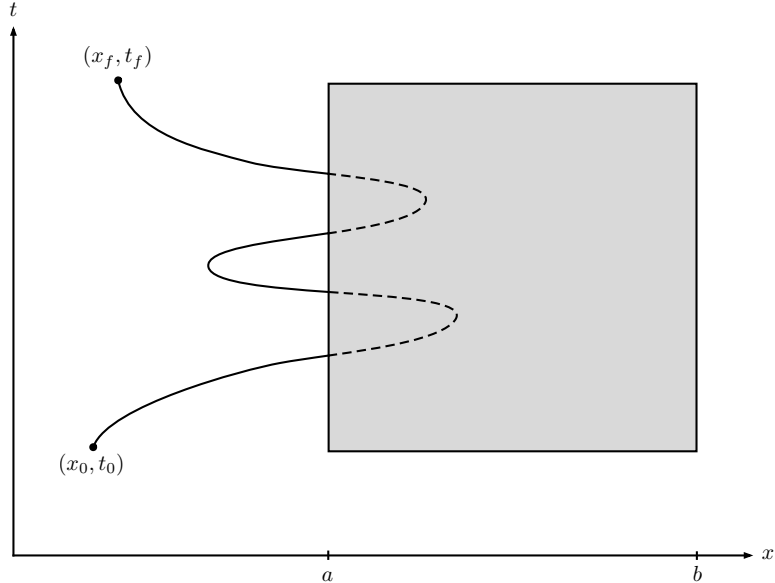}
   \caption{The paths summed over in the modified propagator Eq.(\ref{3.4}). The paths may enter the region $[a,b]$ but their contributions to the sum over paths is exponentially supressed.}
   \label{Fig4}
\end{figure}

Complex potentials of this general type arise in a variety of contexts and have been extensively studied \cite{complex,Hal1,All,PMS}. They will in general still involve reflection, but they behave in the classically expected way, i.e. they are absorbing, for sufficiently small $V_0$,
which is what is required for the propagator to have the intuitively correct properties.
Generalizations of the above scheme are possible in which a real potential of step function form is included. This can help with minimizing reflection and indeed it is possible, for given classes of initial states, to find potentials which are almost perfect absorbers \cite{PMS}.

Given the relationship Eq.(\ref{3.2}), we can now compute the scales associated with reflection, since at least in simple examples, scattering off a complex potential is easy to compute. In the simplest case of a free particle in one dimension scattering off a simple complex step potential $- i V_0 \theta (x)$, reflection is negligible for $V_0 \ll E $, where $E$ is the energy scale of the incoming state \cite{All,HaYe3}.
Because $V_0$ is connected to $\eps$ by Eq.(\ref{3.2a}), this means that the Zeno effect in the string of projectors Eq.(\ref{3.1}) is negligible as long as
\beq
\eps \gg \frac {1} {E}
\label{refcon}
\eeq
Again on general grounds, we expect this to be true in a wide class of models. For more elaborate models in which there are length scales describing the size of the region, the requirement of negligible reflection
may impose further conditions on $E$, in addition to Eq.(\ref{refcon}). However, we assert that for sufficiently large regions $\Delta$, Eq.(\ref{refcon}) is the most important condition.

We thus see that there are two natural and simple ways of defining a restricted propagator outside $\Delta$, in such a way that reflection is minimized but which remain as true as possible to the notion of paths not entering $\Delta$. These definitions involve a new coarse graining parameter $\eps$ which must
in general be chosen to be sufficiently large to avoid unphysical results. The earlier, problematic definitions of the propagators Eqs.(\ref{1.1}), (\ref{1.2}) are obtained in the limit $\eps \rightarrow 0$.

%There may, however, be {\it some} circumstances in which this limit may be taken without problematic features arising, but %this can only be assessed in specific cases \cite{HaWa}.

%[NOTE: This section also needs a generalization of the PDX formula for the case %of
%a complex potential, e.g. as in our arrival time paper.]

Given the modified propagator Eq.(\ref{3.3}) describing restricted propagation outside
$\Delta$, one can also derive a corresponding propagator for entering $\Delta$, analogous
to Eq.(\ref{1.3}) and Eq.(\ref{PDX1}). We define the modified propagator for entering
by
\beq
\hat g^V_{\Delta} (t_f,t_0) = \exp \left( - i H (t_f- t_0) \right)  - \exp \left( - (i H + V_0 P ) (t_f - t_0 ) \right)
\eeq
Some elementary calcuation \cite{HaYe1,Hal} leads to the equivalent form
\beq
\hat g^V_{\Delta} (t_f,t_0) = \int_{t_0}^{t_f} dt \ \exp \left( - i H (t_f- t) \right)
\ V_0 P \  \exp \left( - (i H + V_0 P ) (t - t_0 ) \right)
\label{GV}
\eeq
This may be rewritten
\beq
g^V_\Delta ( \xf, t_f | \x0,t_0) =  V_0 \int_{t_0}^{t_f} dt \int_{\Delta} d^d y
\ g( \xf, t_f | \y, t ) g_r^V (\y, t | \x0,t_0)
\label{GV2}
\eeq
This is the generalization of Eq.(\ref{PDX1}) and tends to it as $V_0 \rightarrow
\infty$. It is different in that firstly, the restricted propagation suppresses paths
that enter $\Delta$ but does not completely exclude them, so the restriction is ``softer''.
Secondly, the intermediate integral is over all of the region $\Delta$, not just
the boundary. With some straightforward calculation in essence the same as a similar case in quantum cosmology considered in Ref.\cite{Hal}, Eq.(\ref{GV}) may be simplified
for small $V_0$ and has the form of an ingoing current operator on the boundary $\Sigma$, the anticipated semiclassical form, and gives intuitively sensible results.

\section{Some Simple Examples}

To illustrate the above ideas, we now give some simple one-dimensional examples.
We consider a free particle in one dimension consisting of mainly positive momenta and initially perfectly localized in $x<0$, so that $ P | \psi \rangle = | \psi \rangle $, where $P = \theta (\hat x)$. We take the spatial region $\Delta$ to be $x>0$ and
consider the following question:
What is the probability that the particle either enters or never enters $\Delta$ during the time interval $[0,\tau]$ and ends in $x_f<0$ at time $\tau$? The situation is depicted in Fig.5.
\begin{figure}[htbp]
   \centering
   \includegraphics[width=4in]{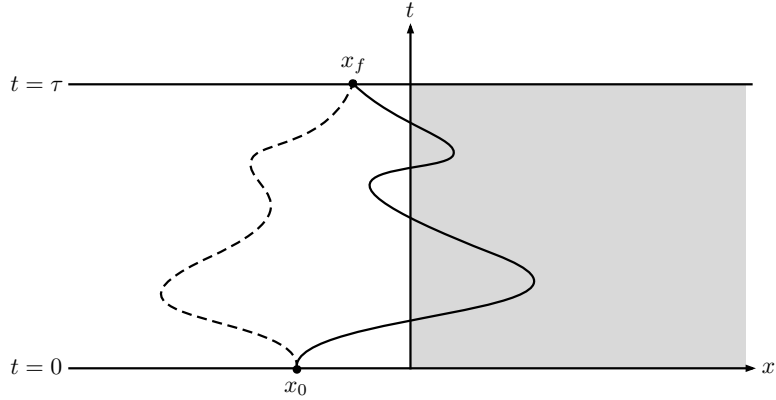}
   \caption{The amplitude for not entering or entering the region $x>0$ during $[0,\tau]$ and ending at any $x_f<0$ at $\tau$ is obtained by summing over paths which respectively, do not enter (dotted line) or enter (bold line) $x>0$.}
   \label{Fig5}
\end{figure}
This question is closely related to the arrival time problem, addressed in many places \cite{time,YDHH}, but here we use it as a simple example of spacetime coarse graining.

The amplitude for not entering $x>0$ is given by the restricted propagator, which in this case is
given by the usual method of images expression
\beq
g_r (x,\tau|x_0,0) = \theta (-x) \theta (-x_0) \left[ g(x,\tau|x_0,0) - g(x,\tau|-x_0,0)\right]
\label{res}
\eeq
and $g$ is the usual free particle propagator
\beq
g(x,\tau|x_0,0) = \left( \frac {m} {2 \pi i \tau} \right)^{1/2} \exp \left( \frac { i m
(x-x_0)^2 } {2\tau} \right)
\eeq
Note the restricted propagator Eq.(\ref{res}) consists of direct and reflected pieces
and hence describes reflection off the origin, as discussed earlier. The restricted propagator is also conveniently written in the operator form
\beq
\hat g_r (\tau, 0) = \bar P (1 - R) e^{ - i H \tau} \bar P
\label{resop}
\eeq
where $\bar P = \theta ( - \hat x)$ and $R$ is the reflection operator,
\beq
R =\int dx | x \rangle \langle - x |
\eeq
Since, as stated in Section 2, $\hat g_r $ is unitary on states with support only in $x<0$ and since the initial state is perfectly localized in $x<0$, the probability for not entering is $p_r =1$.

However, we may still explore the properties of the crossing amplitude to see what sort of result it gives, ignoring the fact that the sum rules will not be satisfied,
as we discussed in the more general case Eq.(\ref{PDX1}).
We therefore consider the propagator $\hat g_\Delta $ for entering $x>0$ during the given time interval, defined by summing over paths which start
at $ x_0 < 0 $ at $t=0$, cross the origin at least once and end $ x <0 $ at $\tau$.
It is most simply expressed in the operator form
\beq
\hat g_\Delta (\tau,0) = \bar P \left (\hat g (\tau,0) - \hat g_r (\tau,0) \right) \bar P
\eeq
where $\hat g = \exp ( - i H \tau) $ is the free particle propagator.
Using Eq.(\ref{resop}), this is equivalently given by
\beq
\hat g_\Delta (\tau,0) = \bar P R e^{ - i H \tau} \bar P
\eeq
(This is in fact equivalent to the PDX form of the crossing propagator, Eq.(\ref{PDX1})).
The probability for entering $\Delta$ is then given by
\bea
p_\Delta (0,\tau) &=& \langle \psi | \hat g_\Delta (\tau,0)^\dag \hat g_\Delta (\tau,0) | \psi \rangle
\nonumber \\
&=& \langle \psi | \bar P P (\tau) \bar P | \psi \rangle
\label{4.7}
\eea
because $ R \bar P R = P = \theta ( \hat x)$. Now note that we may write
\beq
P (\tau ) = P + \int_0^\tau dt \hat J (t)
\eeq
where $ \hat J = \frac {1} {2m} ( \hat p \delta (\hat x) + \delta ( \hat x) \hat p )
$ is the current operator, so we finally have
\beq
p_\Delta (0,\tau) =  \int_0^\tau dt\  \langle \psi | \hat J (t) | \psi \rangle
\label{prob1}
\eeq
This is, on the face of it, a familiar semiclassical answer for the probability for entering $x>0$ during $[0,\tau]$ -- the flux across the origin \cite{time,YDHH}.
It gives a probability close to $1$ for incoming states which substantially cross during the time interval $[0,\tau]$ and probability close to zero for states which substantially miss the interval.

This result is problematic for two reasons. Firstly, because, as stated earlier, the sum rules are generally not obeyed, so that $p_r + p_c \ne 1$ in general. For example, for a state which substantially crosses during the time interval we get $p_r + p_\Delta \approx 2$, a rather striking violation of the sum rules! The sum rules are respected only for states which substantially miss the interval, for which $p_\Delta = 0$.

Secondly, even aside from the failure of the sum rules, the result is misleading. Recall that the histories are required to end in $x<0$ at the final time after entering $x>0$. Semiclassically, such histories would have probability approximately zero for a free particle. Hence Eq.(\ref{prob1}) is semiclassically incorrect for this model.

The point is emphasized by considering
a slightly more complicated version of the above problem.
We take the same initial state in $x<0$ and ask a modified question: What is the probability that the particle either enters or does not enter $x>0$ during the time interval $[0,\tau]$ and ends at any point $x_f$ at time $t_f> \tau $? The situation is depicted in Fig.6.
\begin{figure}[htbp]
   \centering
   \includegraphics[width=4in]{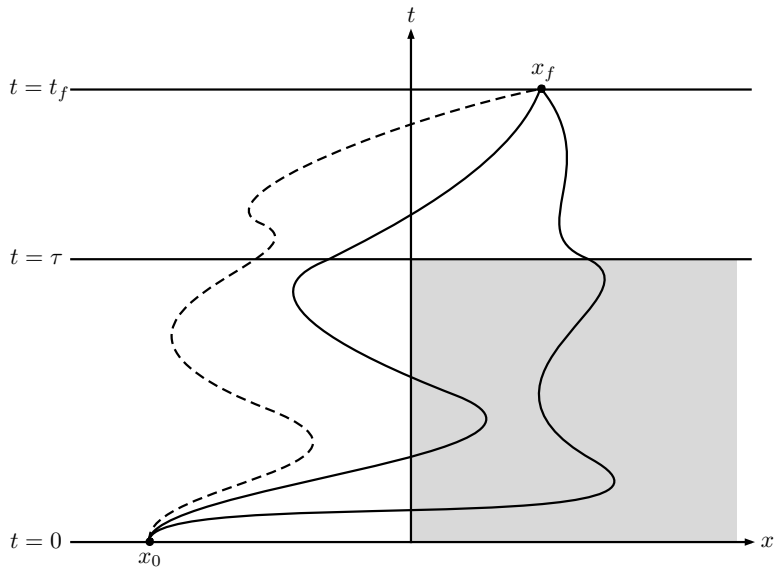}
   \caption{The amplitude for not entering or entering the region $x>0$ during $[0,\tau]$ and ending at any final $x_f$ at $t_f$ is obtained by summing over paths which respectively, do not enter (dotted line) or enter (bold lines) $x>0$ during $[0,\tau]$.}
   \label{Fig6}
\end{figure}
The propagator for not entering is very similar,
\beq
\hat g_r (t_f, 0) = e^{ - i H (t_f - \tau) } \bar P (1 - R) e^{ - i H \tau} \bar P
\label{resop2}
\eeq
so we still have $p_r = 1 $ and again the sum rules are not obeyed.
The propagator for entering, however, acquires a new type of term,
\bea
\hat g_\Delta (t_f,0) &=&  \left( \hat g (t_f,0) - \hat g_r (t_f,0) \right) \bar P
\nonumber \\
&=& e^{ - i H (t_f - \tau) } \left( \bar P R e^{ - i H \tau} \bar P
+ P e^{ - i H \tau} \bar P \right)
\label{4.11}
\eea
because there is now the new possibility that the particle can be in $x>0$ at time $\tau$.
It is easily seen that these two terms make identical contributions to the probability and we therefore have
\beq
p_\Delta (t_f,0) = 2 \int_0^\tau dt\  \langle \psi | \hat J (t) | \psi \rangle
\label{4.12}
\eeq
This is twice the expected semiclassical answer.

The underlying problem is, in essence, the reflection produced by the restricted propagator which persists into the crossing propagator as one can see in Eq.(\ref{4.11}).
As described in Section 3, a solution to this difficulty is obtained by softening the coarse graining by using a complex potential to characterize the restrictions on paths or by using projections not acting at every time. This is described in Refs.\cite{HaYe1,HaYe2}
and we briefly summarize the key ideas here for the complex potential case applied
to the second model considered above.
The amplitude for not entering the region $x>0$ during the time interval $[0,\tau]$ is then given by an expression of the form Eq.(\ref{3.3}), where $P = \theta (\hat x)$. For $V_0 \ll E $, where $E$ is the energy scale of the incoming state, there is negligible reflection, so the part of the state crossing the origin during $[0,\tau]$ will be absorbed and the remainder of the state is unchanged. This effectively means that under propagator with Eq.(\ref{3.3}),
\beq
\hat g_r^V (0,\tau) | \psi \rangle \approx \bar P (\tau ) | \psi \rangle
\eeq
This is the key property that makes the amplitudes defined with a complex potential give sensible physical results. For states consisting of single wave packets reasonably well peaked in position and momentum, the sum rules are satisfied approximately and the crossing probability is approximately Eq.(\ref{prob1}) (with small modifications depending on $V_0$).
The term involving reflection in Eq.(\ref{4.11}) is essentially suppressed which is why the $2$ becomes a $1$ in Eq.(\ref{4.12}).

%These two simple models confirm the general properties of path integral amplitudes discussed in %Sections 2 and 3.

A final related example is that of Yamada and Takagi \cite{YaT}, who took a general initial state and asked for the probability of either crossing or not crossing the origin in either direction during the time interval $[0,\tau]$. The non-crossing propagator is similar to that above, Eq.(\ref{resop}) and is given by
\beq
\hat g_r (\tau, 0) = \bar P (1 - R) e^{ - i H \tau} \bar P +  P (1 - R) e^{ - i H \tau} P
\eeq
They found that the sum rules are satisfied only for states antisymmetric about the origin and that the crossing probability is exactly zero, once more a physical counter-intuitive result.

The analysis of this situation with one of the above modified propagators essentially follows from the work described in Ref.\cite{Ye0}.  For superpositions of incoming wave packets (in either direction), the sum rules are approximately satisfied if the energy scale of the wave packet $E$ satisfies $ E \gg V_0$ and the probabilities are the expected semiclassical ones, so intuitive properties are restored.

Note that this model actually has {\it two} different regimes in which the sum rules are satisfied: approximately for small $V_0$ and exactly, in the Zeno limit of $V_0 \rightarrow \infty$ for antisymmetric initial states, but the resulting probabilities are very different in each case.

Note also that this simple model suggests that the considerations of this paper may have implications for quantum measure theory, in which it is asserted that histories with probability zero do not occur \cite{Sor, Sal}. Intuitively, one would expect a non-zero crossing probability in this model, but the crossing probability is zero in the Zeno limit case. Hence the predictions of quantum measure theory may conflict with the intuitively expected result in this limit.

\section{Connections to Other Work}

The potential difficulties with path integral expressions highlighted in this paper first arose in applications of the decoherent histories approach to spacetime coarse grainings in non-relativistic quantum mechanics \cite{YaT,Yam2,Yam3,Har4,MiHa,HaZa,Har3}. The problems with the Zeno effect did not seem to be appreciated except that Hartle noted that some coarse grainings are ``too strong for decoherence'' \cite{Har4}. The problems also persisted in applications of the decoherent histories approach to quantum cosmology and reparametrization invariant theories \cite{Har3,HaMa,HaTh1,HaTh2,HaWa}. These issue also have consequences for the continuous tensor product structure in the decoherent histories approach, discussed by Isham and collaborators \cite{IsLi,ILSS}.

The specific problem with the Zeno effect was noticed in Ref.\cite{HaWa}. It was also observed that it can in fact be avoided if the spacetime regions involved in the coarse graining are carefully chosen so that the initial state has zero current across the boundary. This resolution has been pursued \cite{ChWa}.

In this paper we have presented a generally applicable resolution
to the Zeno problem involving a softening of the coarse-graining. This was first proposed in the specific context of the arrival time problem in Refs.\cite{HaYe1,HaYe2,Ye4} and in a quantum cosmological model in Ref.\cite{Hal}.

Numerous studies of the dwell and tunneling time problem involve path integral expression of the type considered here (for example, Refs.\cite{Fer,Yam1,GVY,Sok}). However, some of these applications are typically connected to specific models of measurements for which the problems described in this paper may not apply. Indeed, specific models for the measurement of time frequently lead to path integral constructions in which the softening of the coarse graining of the type described in Section 3 is already implemented.
A detailed analysis of the dwell time problem will be given in another publication \cite{HaYe}.

We also stress that partitioning of the paths of the type involved in Eqs.(\ref{1.1}) and (\ref{1.2}) is often used as part of a given calculation (for example, when there is a potential of window function form concentrated in a given region), and the issues raised here are not problematic in such cases.
The issues raised in the present paper concern the possible {\it interpretation} of amplitudes obtained by restricted sums over paths.

%[MORE THOUGHT REQUIRED HERE].

%Is stuff by Yamada about tunneling times and relations between various formula relevant?

In terms of the proposed solutions to the problems with path integrals in Section 3,
our considerations relating to modified coarse grainings have some connection to continuous quantum measurement theory \cite{JaSt,Caves,Mensky,Mensky2}.
We note also that there could be other possible ways of softening the coarse graining
to improve the properties of amplitudes constructed by path integrals.
For example, Marchewka and Schuss have considered path integrals with absorbing boundary conditions \cite{MaSch}.
We also note the interesting and perhaps relevant observations concerning the rigorous definition of path integrals by Sorkin \cite{Sor1}, Geroch \cite{Ger} and Klauder \cite{Kla}.

\section{Summary and Conclusions}

We have argued that amplitudes constructed by path integrals for questions involving time in a non-trivial way can, if implemented in the simplest and most obvious way, lead to problems due to the Zeno effect. This has the consequence that they do not have a sensible classical limit and have properties very different to those expected from the underlying intuitive picture. When path integrals are used in the decoherent histories approach, the Zeno effect can have the consequence that the sum rules are not satisfied except for very trivial initial states.
These problems have been observed in numerous examples and applications but here we have argued that the issue is a very general one to do with the use of path integrals in a wide variety of applications.

We outlined a successful solution to the Zeno problem, through a softening of the coarse graining. Again, this solution has been put forward in specific examples and applications, but here we stress that such a solution will offer a very general solution to the problem. We also note that the softening of the coarse graining introduces, perhaps unexpectedly, one or more new coarse graining parameters, in the simplest case a timescale $\epsilon$, describing the precision to within which the paths are monitored in time. The Zeno problem is then avoided as long as $ \epsilon \gg 1/ E $, where $E$ is the energy scale of the incoming state.

These observations about path integral amplitudes apply most strongly to any approach to quantum theory which involves constructing amplitudes for spacetime coarse grainings which do not refer to a particular measurement scheme, such as the decoherent histories approach and quantum measure theory. Many approaches to quantum gravity are of this type. However, these observations may be less relevant to measurement-based models of spacetime coarse grainings.

We certainly do not claim that any of the papers in this area are wrong. Indeed many authors have noted the unphysical nature of their results. The present paper is, if anything, a cautionary note on the use of path integrals in space time coarse grainings: some types of ``obvious'' coarse-grainings are quite simply unphysical. However,
amplitudes with sensible intuitive properties {\it may} be successfully constructed with proper attention to the implementation of the coarse graining and to the associated time scale.

\section{Acknowledgements}

JMY was supported by the Templeton Foundation. We are grateful to Charis Anastopoulos, Fay Dowker and Larry Schulman for useful comments on an earlier version of the manuscript.

\bibliography{apssamp}

\end{document}